\documentclass[conference]{IEEEtran}
\IEEEoverridecommandlockouts
\usepackage{amsmath,amssymb,amsfonts}
\usepackage{algorithm}
\usepackage{algpseudocode}
\usepackage{graphicx}
\usepackage{textcomp}
\usepackage{booktabs}
\usepackage{array}
\usepackage{xcolor}
\usepackage{makecell}
\usepackage{listings}
\usepackage{multirow}
\lstdefinestyle{mylistingstyle}{
  basicstyle=\ttfamily\footnotesize,     
  breaklines=true,
  aboveskip=15pt, 
  breakindent=2em,
  breakatwhitespace=true,
  frame=lines,                    
  rulecolor=\color{black},         
  belowcaptionskip=1\baselineskip,
  showstringspaces=false,         
  basicstyle=\footnotesize\ttfamily,
}
\usepackage{tikz}
\usepackage{microtype}
\usetikzlibrary{arrows, positioning, shapes.geometric}
\usepackage[numbers]{natbib}
\def\BibTeX{{\rm B\kern-.05em{\sc i\kern-.025em b}\kern-.08em
    T\kern-.1667em\lower.7ex\hbox{E}\kern-.125emX}}
\begin{document}

\title{Distributed Log-driven Anomaly Detection System based on Evolving Decision Making}

\author{
Zhuoran Tan*, Qiyuan Wang*, Christos Anagnostopoulos*, Shameem P. Parambath*, Jeremy Singer*, Sam Temple†  
\\  
*School of Computing Science, University of Glasgow, Glasgow, UK  
\\  
†JUMPSEC Ltd, UK  
\\  
\{z.tan.1, Qiyuan.Wang, Christos.Anagnostopoulos, Sham.Puthiya, jeremy.singer\}@glasgow.ac.uk\\sam.temple@jumpsec.com  
}

\maketitle

\begin{abstract}
Effective anomaly detection from logs is crucial for enhancing cybersecurity defenses by enabling the early identification of threats. Despite advances in anomaly detection, existing systems often fall short in areas such as post-detection validation, scalability, and effective maintenance. These limitations not only hinder the detection of new threats but also impair overall system performance.
To address these challenges, we propose CEDLog, a novel practical framework that integrates Elastic Weight Consolidation (EWC) for continual learning and implements distributed computing for scalable processing by integrating Apache Airflow and Dask.
In CEDLog, anomalies are detected through the synthesis of Multi-layer Perceptron (MLP) and Graph Convolutional Networks (GCNs) using critical features present in event logs. Through comparisons with update strategies on large-scale datasets, we demonstrate the strengths of CEDLog, showcasing efficient updates and low false positives.
\end{abstract}


\section{Introduction}
Logs are a key data source for anomaly detection, helping to mitigate cyber threats. Recent methods range from Machine Learning (ML)\cite{zhangLogPromptLogbasedAnomaly2023, catilloAutoLogAnomalyDetection2022a} to provenance graph-based analysis \cite{niu_logtracer_2022, li_t-trace_2024}, typically involving log parsing, feature generation, and anomaly detection. 
Benchmarks on public datasets like BlueGene/L (BGL) and Hadoop Distributed File System (HDFS) \cite{zhuLoghubLargeCollection2023}  highlight high precision yet reveal notable challenges in post-detection validation for sequential (streaming) data. 
Post-detection validation refers to the process of verifying and refining the results of a detection system after an initial identification has been made \cite{6513558}. 
While some approaches integrate feedback mechanisms\cite{zhangLogPromptLogbasedAnomaly2023, 10.1145/3133956.3134015}, they lack strategies to incorporate feedback without degrading prior learning.

This paper introduces CEDLog, a scalable log anomaly detection framework designed for scalability and reliable maintenance. It features parallel processing and distributed computing leveraging Apache Airflow\footnote{https://Airflow.apache.org/} and Dask\footnote{https://www.dask.org/} to enhance efficiency, especially for structured log data. To handle format variability, it integrates Elasticsearch, Logstash, and Kibana (ELK) stack\footnote{https://www.elastic.co/elastic-stack}, complemented with a Human-in-the-Loop (HITL) mechanism for adaptive feedback. 

CEDLog combines distributed computing with evolving decision-making. Airflow and Dask enable scalable execution, while continual learning with EWC mitigates catastrophic forgetting \cite{kirkpatrickOvercomingCatastrophicForgetting2017}. 
CEDLog is deployed using Docker\footnote{https://www.docker.com/} for offline training and online inference, enabling real-time log processing with HITL validation.

\section{Related Work}

In this section, we address fundamental issues and review the related literature on log anomaly detection.

\subsection{Preliminaries \& Problem Fundamentals}

Logs are a data key source for threat and anomaly detection, leveraging methods such as signature analysis, pattern recognition, and machine learning.
Based on their purpose, anomalies and threats can be categorized into two main types:

\paragraph{\textbf{Single-Point Anomaly}} A single-point anomaly is an isolated data point in a time series that significantly deviate from the expected pattern. Given a time series:
$$
X = \{x_1, x_2, \dots, x_T\}
$$

where $x_t$ represents the value at time $t$, a single-point anomaly occurs when
$
|x_t - \mu_t| > \lambda\sigma_t
$, where $\mu_t$ is the expected value, $\sigma_t$ is the standard deviation, $\lambda$ is a threshold. 
The single-point anomaly contains failure anomaly, which may arise from exception events, abnormal packet size, and malicious IP addresses or ports. 
The problem to detect single-point anomaly can be framed as either binary or multi-class classification problem.

\paragraph{\textbf{Sequential Anomalies}} A sequential anomaly (or collective anomaly) occurs when a subsequence of values deviates from the normal pattern, even if individual points may not be outliers. For a subsequence:
$$
S_t^k = {x_t, x_{t+1}, \dots, x_{t+k}}
$$
where k is the length of the sequence, an anomaly is detected normally with distance like Euclidean Distance \cite{danielsson1980euclidean}, if:
$$
d(S_t^k, S_{\text{ref}}) = \sum_{i=0}^{k-1} (x_{t+i} - x_{\text{ref},i}) ^ 2
$$

While some studies use sliding windows to detect sequential anomalies, this study focuses on single-point anomalies using binary classification. It detects failure based on benchmark dataset \cite{zhuLoghubLargeCollection2023}, including semi-structured BGL and HDFS logs.



\paragraph{\textbf{Model Deployment}} Apache Airflow is the chosen tool for deploying the proposed detection framework. Unlike other tools like MLflow\footnote{https://mlflow.org/}, Airflow is widely used for extract, transform, and load (ETL) processes, as well as data and ML pipelines. As a workflow orchestration tool, it allows developers to programmatically author, schedule, and monitor workflows as Directed Acyclic Graphs (DAGs). With its Python-based interface, Airflow facilitates the definition of complex workflows, enabling tasks to run sequentially or in parallel while managing dependencies. We leverage this capability to construct sequential, dependent pipelines. Furthermore, its support for dynamic pipeline generation provides flexibility in defining diverse process logic tailored to different clients.

\subsection{Anomaly Detection in Logs}

    Several works have been proposed for failure anomaly detection from logs.
    \citet{catilloAutoLogAnomalyDetection2022a} used auto-encoding to establish a baseline from normal operations.
    \citet{zhangLogPromptLogbasedAnomaly2023} leveraged Pre-trained Language Models (PLMs) with semantic and sequential tokens for anomaly detection. 
    \citet{huangTransferLogbasedAnomaly2020} introduced a framework that trained on labeled source samples while incorporating semantic information. However, these approaches overlook practical challenges such as model maintenance, deployment, and resource requirements \cite{zhangLogPromptLogbasedAnomaly2023}. 

    Some works have explored deployment strategy.
    \citet{liSwissLogRobustAnomaly2023} discussed offline training and online detection, while \citet{skopikBehaviorBasedAnomalyDetection2023} introduced continual learning to integrate the latest data. 
    However, none provided an efficient update strategy to ensure model evolution.
     
    In contrast, CEDLog integrates both the training and inference, employing an online learning mechanism with EWC enhancement to mitigate catastrophic forgetting. 
    It adapts to diverse log types and input features, improving robustness. Optimized for CPU efficiency, CEDLog is well-suitable for small to medium-sized enterprises without relying on high-performance computing like in \citet{zhangLogPromptLogbasedAnomaly2023}.

\section{Methodology}

\begin{figure*}[ht]
    \centering
    \includegraphics[scale=0.55]{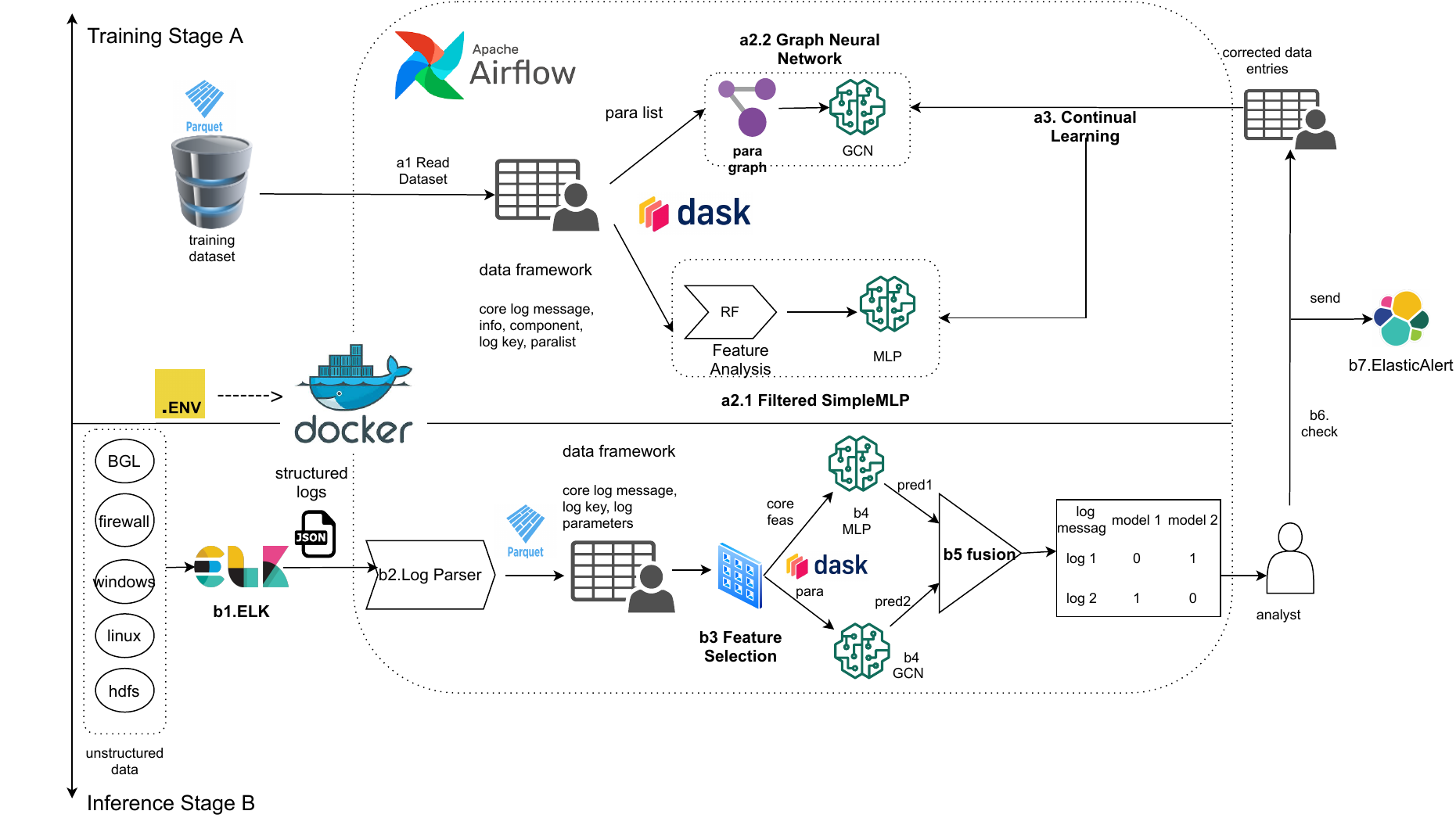}
    \caption{The CEDLog Framework. 
    \textbf{Training Stage:} a1. Read dataset with parquet format; a2.1. Analyse the features and send top weighted features to the MLP; a2.2. Create a graph based on ParameterList, train with GCN; a3. Integrate the correct predicted framework to update both models. \textbf{Inference Stage}: b1. Log transformation with ELK and output JSON format; b2. Log parsing and output data-frame with parquet format; b3. Choose top weighted features; b4. Separately detect with MLP for core features without Parameterlist, and GCN for Parameterlist; b5. Fuse the predicted results; b6 Analyst checks the predicted results; b7 Send the corrected result to ElasticAlert.}
    \label{fig: framework}
\end{figure*}

    To develop a scalable log anomaly detection system capable of addressing diverse threats, we leverage ELK for multi-source log integration. As shown in Figure \ref{fig: framework}, log parser parses transformed structured logs processed by ELK into tabular format. 
    The structured logs undergo anomaly detection within a dedicated engine consisting of multiple processing pipelines. 
    These pipelines are orchestrated using Airflow, where each is defined as an operator within a DAG. The DAG organizes tasks based on their dependencies, ensuring efficient execution. Airflow is deployed in a distributed mode, with Dask integrated for parallel log parsing.
    Finally, detected anomalies are forwarded to ElasticAlert \footnote{https://github.com/jertel/elastalert2}, which generates alerts to clients. 
    The subsequent sections provide in-depth discussions on key components, including log parsing, feature generation, scalable processing, detection model design, maintenance mechanisms, and the final distributed deployment.

\subsection{Log Parsing}

    The log parser serves as the core engine for extracting distinct components and converting raw logs into tabular-format. The parsed structured output makes it more convenient to extract features required for machine learning models. To achieve both high accuracy and exceptional parsing speed, we selected Drain \cite{He2017DrainAO} as the log parser due to its overall performance \cite{zhuLoghubLargeCollection2023} in structuring various raw logs. 

    Drain is a hierarchical clustering-based log parser that achieves efficient log parsing with a fixed depth tree structure. The principle behind Drain can be explained as:

    Each raw log message can be represented as a sequence of tokens:
    $$
    l_i = (w_1, w_2, ..., w_m)
    $$
    A log entry $l_i$ is matched by traversing the tree level by level, using the first few tokens as tree nodes.
    At each depth $d$, the corresponding token $w_d$ is used to traverse the tree:
    $$
    N_d = f(w_d, N_{d-1})
    $$
    where $f(w_d, N_{d-1})$ is the function that selects or creates the next node based on token $w_d$ and the previous node $N_{d-1}$.

    Once a leaf node is reached, the existing templates are checked for a match based on the following similarity score:

    $$
    \text{Similarity}(l_i, T_k) = \frac{\sum\limits_{j=1}^{m} \mathbb{I}(t_j= w_j  \text{ or } t_j = \langle * \rangle)}{m}
    $$

    where $\mathbb{1}$ is an indicator function that counts matching tokens. The equation $t_j= w_j$ points at the fixed token in a log template $T_k$, and $t_j = \langle * \rangle)$ is a wildcard token representing dynamic word. 
    If the similarity exceeds a threshold $\theta$, the log message is assigned to $T_k$. Otherwise, a new template is created. The computational complexity of Drain is $O(D)$, where $D$ is the tree depth and does not grow with log size.

    The final output of log parsing is to extract certain patterns, individual components, and the parameters from original semi-structured logs.  
    Each log entry $l_i$ generally consists of the following attributes:
    $$
    l_i = \left(\begin{aligned}Datetime_i, Context_i, EventTemplate_i, \\  RecordID_i, Log\_Level_i, ParameterList_i\end{aligned}
    \right)
    $$
    
    When parsing on HDFS and BGL, we observe that the extracted dynamic tokens (variables) often contain a mix of contextual and numeric data. This insights drives us to implement fine-grained feature engineering tailored to the structured logs.

    \subsection{Feature Engineering}
    The structured output includes various components, but the factors determining log entry labels and their influence remain unclear. To address this, we use the Random Forest algorithm \cite{8074494} to compute feature importance scores. Well-suited for non-linear relationships, Random Forest leverages information gain via entropy to identify key features influencing labels.  During analysis, variables in \texttt{ParameterList} are concatenated as a single string. The process follows:
    $$
    I = RandomForestFeatureImportance(L)
    $$
    in which $L$ is the dataset. Then a threshold $\tau$ is defined to filter out importance columns $\mathcal{C}$:
    $$
    \mathcal{C} = \{c | c \in columns(L), I(c) \succ \tau\}
    $$
    A weight dictionary $W$ is constructed to map each column to its feature importance:
    $$
    W = \{column: I(column) | column \in \mathcal{C}\}
    $$
    This dictionary is then used when fusing the final classification result.


    To generate a suitable format for the detection model, tailored feature engineering methods are required for different features. Due to dual-model setup, as shown in Fig. \ref{fig: framework}, two groups of input are created: one for feature matrix input and another for graph representation input.

    The first part feature is the feature matrix $X$ excluding \texttt{ParameterList} column, due to `list' type of variables.
    $$
    X = L[\mathcal{C} \setminus \{"ParameterList"\}]
    $$
    This part emphasizes the anomaly from specific event templates, along with other features except the \texttt{ParameterList}.
    
    The second part feature is created by taking the variables inside \texttt{ParameterList} as leaf node and enriching the graph representation with \texttt{Event\_Id} as the root node for every log entry. The graph edges can be represented as:
    $$
    E = \{(e_i, p_{ij}) | p_{ij} \in L_i."ParameterList"\}
    $$
    meaning the each $e_i$ (EventId) connects to every $p_{ij}$ (parameter) in the corresponding \texttt{ParameterList}.
    This part emphasizes the anomaly from abnormal variable values, like extreme packet sizes, in event templates.

    During graph construction, we exclude variables that lack semantic significance, such as block IDs in HDFS logs. For variables like paths, we retain only the last two levels to reduce the impract of personal directory structures. We formulate this task as a binary classification problem at subgraph level.

    After token extraction, the spaCy\footnote{https://spacy.io/} library embeds tokens from node values into numeric vectors of uniform length using pre-trained word embeddings, such as GloVe\footnote{https://nlp.stanford.edu/projects/glove/}. This embedding enhances semantic learning in graph neural training beyond structural patterns, improving classification interpretability through internal semantic similarity.
    
    \subsection{Scalable Processing}

    Optimizing feature generation significantly reduces processing time, accelerating the entire detection task. Feature engineering is often the most time-consuming stage in the ML lifecycle \cite{zhouMLOpsCaseStudy2020}. To enhance efficiency, we apply task-based parallelism, which divides large tasks into smaller, independent ones that run concurrently \cite{9062721}.

    We incorporate Dask's \textbf{map\_partitions} function \cite{Rocklin2015DaskPC} into feature generation, leveraging multi-core processing.
    This function is used for:
    \begin{itemize}
        \item constructing graph representation from log entries
        \item parsing semi-structure logs into individual components
    \end{itemize}
    This implementation significantly improves parsing efficiency, particularly for datasets with millions of log entries.

\subsection{Dual-Model Detection}

    The detection component comprises dual models and post-stage validation that focuses on continuous learning, as shown in Fig. \ref{fig: framework}. The applied MLP model primarily captures the error anomalies from specific event templates and converted numeric log info levels and components. For example, one individual event template in BGL logs is marked as anomaly when appearing in the same when component is equal 'APP'. This specific event template is benign in normal situation without other information. 
    The network architecture of this MLP model is composed of two hidden layers with node units of 64 and 32, following the typical MLP structure.
    Additionally, we include a Batch-Normalization layer after each hidden layer to normalize the scale of encoded strings. 

    The applied GCN model identifies error anomalies within the variables in \texttt{ParameterList(ParaList)}. Especially, errors can arise from extreme values, such as unusually large package sizes, unfamiliar IP addresses, or unexpected file sequences.
    This GCN model draws inspiration from the works of \cite{montiFakeNewsDetection2019a} and \cite{kipfSemiSupervisedClassificationGraph2017a}. The implemented GCN model follows basic architecture to reduce complexity, comprising two sequential graph convolutional layers, one mean pooling layer, two fully connected sequential layers, and one final output layer. 
    The graph convolutional layer, as described in \cite{kipfSemiSupervisedClassificationGraph2017a}, has 64 units and uses ReLu activation. The pooling layer summarizes the node representation learned and creates a graph representation. This representation becomes the input for the two subsequent fully connected layers with 32 and 16 units, respectively. The final output layer contains a single unit for two classes, with \texttt{Sigmoid} as the activation function.
    
    The final result integrates the predictions from both models by considering the portion of input features weight compared with total weight.
    To explain it, the relative importance scores are computed first, which consists of two parts:
    $$
    s_0 = \frac{W[\mathcal{C} \setminus \{"ParaList"\}]}{W_{sum}}, s_1 = \frac{W["ParaList"]}{W_{sum}}
    $$
    The final fused anomaly score $F$ is computed using probability estimates like:
    $$
    F = P(p_1 = 0)* s_0 + P(p_2 = 0) * s_1
    $$
    The final decision rule follows the result of comparsion:
    $$
    \hat{y} = \begin{cases} 
    0, & \text{if } F > 0.5 \\
    1, & \text{otherwise}
    \end{cases}
    $$
    The fusion can generate robust prediction result by considering feature difference and model ability. 
        
\subsection{Human-in-the-Loop Continual Learning}

    We introduce a mechanism for updating the models consistently. 
    During manual inspection, if an analyst identifies a FP prediction, this signals an error in the prediction of the GCN model, contributing to an incorrect prediction. The misclassified event is then flagged as false and added to the next round as fine-tuning data. These flagged data contribute to the model update while the updated DAG is triggered at specific intervals.
    
    During the update process, we integrate elastic weight consolidation (EWC) \cite{Kutalev2021StabilizingEW}, as a method to prevent catastrophic forgetting, into the update pipelines.
    The core principle of EWC is to penalize changes to important parameters of previous tasks using a quadratic regularization term based on the Fisher Information Matrix \cite{karakida2019universal}. The loss function is represented as:
    $$
    \mathcal{L}_{\text{EWC}} (\theta) = \mathcal{L}_(\theta) + \frac{\lambda}{2}\sum_i F_i (\theta_i - \theta_i^*)^2
    $$
    where $\mathcal{L}_(\theta)$ is the standard loss function for the current task. $\theta^*$ are the optimal parameters learned from previous tasks. $\lambda$ is a hyperparameter that controls the strength of the regularization. $F_i$ is the Fisher Information Matrix, which estimates the importance of each parameter $\theta_i$ for previous tasks.

\subsection{Scalable Deployment}

    To achieve scalability, particularly with multiple logs for individual clients, configuring Airflow in distributed way is crucial for improving availability and scalability. The Celery Executor\footnote{https://airflow.apache.org/docs/apache-airflow-providers-
    celery/stable/celery executor.html} is chosen to deploy in a distributed setup. Celery Executor is based on Python Celery\footnote{https://github.com/celery/celery} to process asynchronous tasks, which is designed for distributed environments and can distribute tasks across multiple nodes, enhancing scalability. 

\begin{figure}[ht]
    \centering
    \includegraphics[scale=0.63]{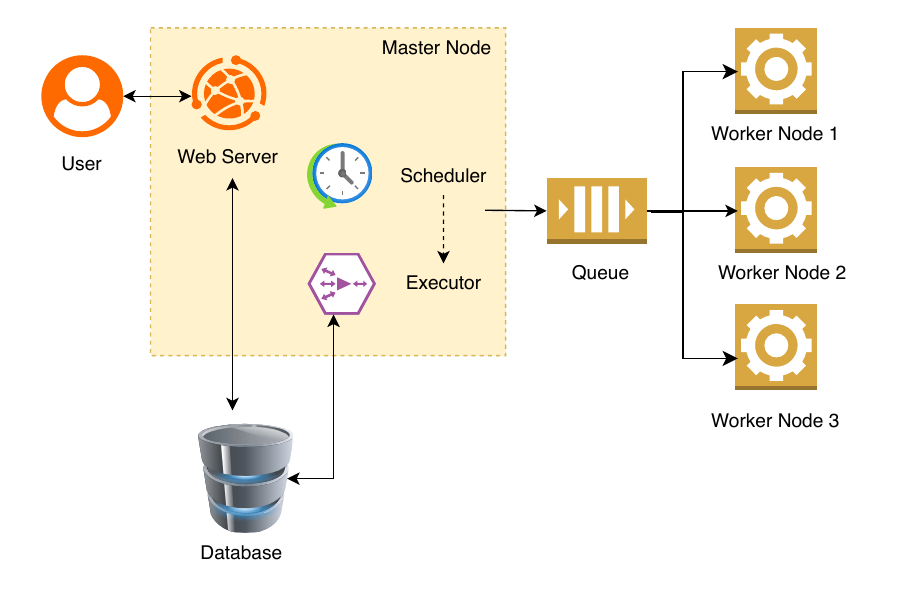}
    \caption{Airflow in Distributed Mode}
    \label{fig: distributed airflow}
    \end{figure}

    The Scheduler in CeleryExecutor adds all tasks to the task queue as shown in Figure \ref{fig: distributed airflow}. Celery workers pull tasks from the queue and execute them. After the execution is completed, the worker reports the status of the task in the database. The Scheduler knows from the database when a task has been completed and then runs the next set of tasks or process alerts based on what is configured in the DAG.

\section{Evaluation}

    The environment to evaluate the performance is a Debian GNU/Linux 10 (buster) node, with 12 CPU cores,  64GB of memory and 100GB disk space. The training size of 0.8 for all evaluations. 
    The datasets refer to \cite{zhuLoghubLargeCollection2023}, in which two labeled data, including HDFS and BGL, are widely adopted.
    
    The chosen HDFS dataset is the second version, simulated in a cloud environment using benchmark workloads. It contains 10 million logs collected over 39 hours, involving one name node and 32 data nodes. 
    The BGL logs were collected through a supercomputer system in 214 days, with a total size of around 5 million labeled logs. The dataset contains alert and non-alert messages identified by alert category tags.
    During the following evaluation, 2 million HDFS logs and 2.5 million BGL logs are chosen due to computational resource limitations.

\subsection{Fusion Decision Making}

       \begin{table}[ht]
        \centering
        \footnotesize
        \caption{Performance with Decision Fusion}
        \begin{tabular}{l|cccccc}
        \hline
        \textbf{Dataset}      & \textbf{Model} & \textbf{Acc} & \textbf{Prec} & \textbf{F1-score} & \textbf{Recall} & \textbf{FPR} \\ \hline
        \multirow{3}{*}{HDFS} & MLP         &  0.9747            &   0.9585        &      0.0362          &        0.0185        &  0.00002   \\ \cline{2-7} 
                              & GCN         &  0.9731           &  0.8983            &    0.169            &     0.0933            &  0.0003     \\ \cline{2-7} 
                              & Fusion         &    0.9715          &        1.0              &       0.0517            &     0.0265 &       0.0\\ \hline
        \multirow{3}{*}{BGL} & MLP         &    0.9698      &   0.8356           &    0.7815          &  0.7341         &  0.0115      \\ \cline{2-7} 
                              & GCN         &    0.9091    &    0.7369           &    0.0625           &   0.0327           &   0.0012        \\ \cline{2-7} 
                              & Fusion         &    0.9890          &   0.9196              &    0.9422        &    0.9659       &    0.0086        \\ \hline
        \end{tabular}
        
        \label{fig: decision fusion}
        \end{table}
    

    
    An enhancement involves the decision making fusion of the predictions from MLP and GCN. 
    The fusion decision balances the likelihood of predicted labels with the importance percentage of input features.
    This method emphasizes the factors of the core information during the decision-making process. 
    
    As shown in Table \ref{fig: decision fusion}, the precision metric of the \texttt{Fusion} model is higher in BGL logs, with a value of 1, compared to 0.9585 for MLP and 0.8983 for GCN. 
    In HDFS logs, the precision improvement with the \texttt{Fusion} model is even more pronounced, achieving 0.9196, compared to 0.8356 for MLP and 0.7369 for GCN. Additionally, the False Positive Rate (FPR) of the \texttt{Fusion} model is close to zero in both logs.
    Furthermore, the degradation of accuracy is negligible.
    In the BGL logs as shown in Table \ref{fig: decision fusion}, the improvements in both precision and accuracy are expected, demonstrating the robustness of the prediction. The FPR value potentially balances the performance of the two models, which is also an acceptable low value.

\subsection{Continual Learning with EWC}

    \begin{table}[ht]
    \caption{Performance with EWC on MLP for BGL Logs}
    \label{tab:continuallearning}
    \resizebox{\columnwidth}{!}{\begin{tabular}{l|cccccc}
    \hline
    \textbf{Train Type}              & \textbf{Task} & \textbf{Acc} & \textbf{Prec} & \textbf{F1-Score} & \textbf{Recall} & \textbf{FPR} \\ \hline
    Initial Train                     & A  &  0.985   &  0.9719    &    0.9718       &   0.9717    &  0.0283   \\ \hline
    \multirow{2}{*}{Norm Retrain}   & A  &  0.9876   &  0.9827    &      0.9766    &    0.9705    &   0.0062  \\ \cline{2-7} 
                                        & B  &  0.9876   & 0.9825     &   0.9764       &   0.9704     &   0.0062
                                        \\ \hline
    \multirow{2}{*}{EWC Retrain} & A  &  0.9888   &   0.9874   &    0.9789      &   0.9705     &  0.0045  \\ \cline{2-7} 
                                & B  &  0.9892   & 0.9886     &      0.9794    &   0.9704    &   0.0296
                                         \\ \hline
    \end{tabular}}
    \end{table}

    As illustrated in Table \ref{tab:continuallearning}, we present a performance comparison of various retraining methods for the MLP model using BGL logs. 
    The initial performance evaluations are performed using the same dataset as in the initial task \textit{A}. 
    The task \textit{B} refers to a new prediction task on a new dataset. 
    By comparing training strategies, the accuracy slightly improves from 0.985 (initial) to 0.9852 (EWC Retrain). Additionally, precision and F1-score also improve with EWC compared to normal retraining. Recall remains almost the same across all models.
    The FPR is lowest in EWC for Task A (0.0045) but slightly higher for Task B (0.0296).

    We can verify that EWC maintains highly accuracy across different tasks while reducing the catestrophic forgetting problem seen in normal retraining.
    Lower FPR in task A for EWC suggests that it improves robustness in distinguishing normal and anomalous logs.
    EWC retrain increases the FPR in task B, which may indicate a challenge in balancing knowledge retention across tasks.

\section{Conclusions}

    We introduce CEDLog, a distributed, continually evolving framework for log anomaly detection. It utilizes Airflow for distributed deployment and Dask for parallel processing, enabling efficient large-scale detection. CEDLog fuses results from dual models, each targeting distinct feature groups, and employs continual learning with EWC to ensure consistent update without degrading performance. It achieves high precision and a low false positive rate. To support multiple clients, we plan to integrate Kubernetes\footnote{https://kubernetes.io/} for synchronous monitoring and. implement comprehensive attack simulations, such as red teaming, to evaluate detection capability across diverse scenarios.
    
\section*{Acknowledgments}

    This work has received technical support from colleagues at JUMPSEC Ltd in testing and validating the developed infrastructure.



\begin{thebibliography}{21}
\providecommand{\natexlab}[1]{#1}
\providecommand{\url}[1]{#1}
\csname url@samestyle\endcsname
\providecommand{\newblock}{\relax}
\providecommand{\bibinfo}[2]{#2}
\providecommand{\BIBentrySTDinterwordspacing}{\spaceskip=0pt\relax}
\providecommand{\BIBentryALTinterwordstretchfactor}{4}
\providecommand{\BIBentryALTinterwordspacing}{\spaceskip=\fontdimen2\font plus
\BIBentryALTinterwordstretchfactor\fontdimen3\font minus \fontdimen4\font\relax}
\providecommand{\BIBforeignlanguage}[2]{{%
\expandafter\ifx\csname l@#1\endcsname\relax
\typeout{** WARNING: IEEEtranN.bst: No hyphenation pattern has been}%
\typeout{** loaded for the language `#1'. Using the pattern for}%
\typeout{** the default language instead.}%
\else
\language=\csname l@#1\endcsname
\fi
#2}}
\providecommand{\BIBdecl}{\relax}
\BIBdecl

\bibitem[Zhang et~al.(2023)Zhang, Huang, Zhao, Bian, and Du]{zhangLogPromptLogbasedAnomaly2023}
T.~Zhang, X.~Huang, W.~Zhao, S.~Bian, and P.~Du, ``{{LogPrompt}}: {{A Log-based Anomaly Detection Framework Using Prompts}},'' in \emph{2023 {{International Joint Conference}} on {{Neural Networks}} ({{IJCNN}})}.\hskip 1em plus 0.5em minus 0.4em\relax {Gold Coast, Australia}: {IEEE}, 2023, pp. 1--8.

\bibitem[Catillo et~al.(2022)Catillo, Pecchia, and Villano]{catilloAutoLogAnomalyDetection2022a}
M.~Catillo, A.~Pecchia, and U.~Villano, ``{{AutoLog}}: {{Anomaly}} detection by deep autoencoding of system logs,'' \emph{Expert Systems with Applications}, vol. 191, p. 116263, 2022.

\bibitem[Niu et~al.(2022)Niu, Yu, Li, Li, Zhang, and Zhang]{niu_logtracer_2022}
W.~Niu, Z.~Yu, Z.~Li, B.~Li, R.~Zhang, and X.~Zhang, ``{LogTracer}: {Efficient} {Anomaly} {Tracing} {Combining} {System} {Log} {Detection} and {Provenance} {Graph},'' in \emph{{GLOBECOM} 2022 - 2022 {IEEE} {Global} {Communications} {Conference}}.\hskip 1em plus 0.5em minus 0.4em\relax IEEE, Dec. 2022, pp. 3356--3361.

\bibitem[Li et~al.(2024)Li, Liu, Qiao, Zhu, Shen, and Ma]{li_t-trace_2024}
T.~Li, X.~Liu, W.~Qiao, X.~Zhu, Y.~Shen, and J.~Ma, ``\BIBforeignlanguage{en}{T-{Trace}: {Constructing} the {APTs} {Provenance} {Graphs} {Through} {Multiple} {Syslogs} {Correlation}},'' \emph{\BIBforeignlanguage{en}{IEEE Transactions on Dependable and Secure Computing}}, vol.~21, no.~3, pp. 1179--1195, May 2024.

\bibitem[Zhu et~al.(2023)Zhu, He, He, Liu, and Lyu]{zhuLoghubLargeCollection2023}
J.~Zhu, S.~He, P.~He, J.~Liu, and M.~R. Lyu, ``Loghub: {{A Large Collection}} of {{System Log Datasets}} for {{AI-driven Log Analytics}},'' 2023.

\bibitem[DeOrio et~al.(2013)DeOrio, Li, Burgess, and Bertacco]{6513558}
A.~DeOrio, Q.~Li, M.~Burgess, and V.~Bertacco, ``Machine learning-based anomaly detection for post-silicon bug diagnosis,'' in \emph{2013 Design, Automation \& Test in Europe Conference \& Exhibition (DATE)}, 2013, pp. 491--496.

\bibitem[Du et~al.(2017)Du, Li, Zheng, and Srikumar]{10.1145/3133956.3134015}
M.~Du, F.~Li, G.~Zheng, and V.~Srikumar, ``Deeplog: Anomaly detection and diagnosis from system logs through deep learning,'' in \emph{Proceedings of the 2017 ACM SIGSAC Conference on Computer and Communications Security}, ser. CCS '17.\hskip 1em plus 0.5em minus 0.4em\relax New York, NY, USA: Association for Computing Machinery, 2017, p. 1285–1298.

\bibitem[Kirkpatrick et~al.(2017)Kirkpatrick, Pascanu, Rabinowitz, Veness, Desjardins, Rusu, Milan, Quan, Ramalho, {Grabska-Barwinska}, Hassabis, Clopath, Kumaran, and Hadsell]{kirkpatrickOvercomingCatastrophicForgetting2017}
J.~Kirkpatrick, R.~Pascanu, N.~Rabinowitz, J.~Veness, G.~Desjardins, A.~A. Rusu, K.~Milan, J.~Quan, T.~Ramalho, A.~{Grabska-Barwinska}, D.~Hassabis, C.~Clopath, D.~Kumaran, and R.~Hadsell, ``Overcoming catastrophic forgetting in neural networks,'' \emph{Proceedings of the National Academy of Sciences}, vol. 114, no.~13, pp. 3521--3526, 2017.

\bibitem[Danielsson(1980)]{danielsson1980euclidean}
P.-E. Danielsson, ``Euclidean distance mapping,'' \emph{Computer Graphics and image processing}, vol.~14, no.~3, pp. 227--248, 1980.

\bibitem[Huang et~al.(2020)Huang, Liu, Fung, He, Zhao, Yang, and Luan]{huangTransferLogbasedAnomaly2020}
S.~Huang, Y.~Liu, C.~Fung, R.~He, Y.~Zhao, H.~Yang, and Z.~Luan, ``Transfer {{Log-based Anomaly Detection}} with {{Pseudo Labels}},'' in \emph{2020 16th {{International Conference}} on {{Network}} and {{Service Management}} ({{CNSM}})}.\hskip 1em plus 0.5em minus 0.4em\relax {Izmir, Turkey}: {IEEE}, 2020, pp. 1--5.

\bibitem[Li et~al.(2023)Li, Chen, Jing, He, and Yu]{liSwissLogRobustAnomaly2023}
X.~Li, P.~Chen, L.~Jing, Z.~He, and G.~Yu, ``{{SwissLog}}: {{Robust Anomaly Detection}} and {{Localization}} for {{Interleaved Unstructured Logs}},'' \emph{IEEE Transactions on Dependable and Secure Computing}, vol.~20, no.~4, pp. 2762--2780, 2023.

\bibitem[Skopik et~al.(2023)Skopik, Wurzenberger, H{\"o}ld, Landauer, and Kuhn]{skopikBehaviorBasedAnomalyDetection2023}
F.~Skopik, M.~Wurzenberger, G.~H{\"o}ld, M.~Landauer, and W.~Kuhn, ``Behavior-{{Based Anomaly Detection}} in {{Log Data}} of {{Physical Access Control Systems}},'' \emph{IEEE Transactions on Dependable and Secure Computing}, vol.~20, no.~4, pp. 3158--3175, 2023.

\bibitem[He et~al.(2017)He, Zhu, Zheng, and Lyu]{He2017DrainAO}
P.~He, J.~Zhu, Z.~Zheng, and M.~R. Lyu, ``Drain: An online log parsing approach with fixed depth tree,'' \emph{2017 IEEE International Conference on Web Services (ICWS)}, pp. 33--40, 2017.

\bibitem[Jaiswal and Samikannu(2017)]{8074494}
J.~K. Jaiswal and R.~Samikannu, ``Application of random forest algorithm on feature subset selection and classification and regression,'' in \emph{2017 World Congress on Computing and Communication Technologies (WCCCT)}, 2017, pp. 65--68.

\bibitem[Zhou et~al.(2020)Zhou, Yu, and Ding]{zhouMLOpsCaseStudy2020}
Y.~Zhou, Y.~Yu, and B.~Ding, ``Towards {{MLOps}}: {{A Case Study}} of {{ML Pipeline Platform}},'' in \emph{2020 {{International Conference}} on {{Artificial Intelligence}} and {{Computer Engineering}} ({{ICAICE}})}.\hskip 1em plus 0.5em minus 0.4em\relax {Beijing, China}: {IEEE}, 2020, pp. 494--500.

\bibitem[Slaughter and Aiken(2019)]{9062721}
E.~Slaughter and A.~Aiken, ``Pygion: Flexible, scalable task-based parallelism with python,'' in \emph{2019 IEEE/ACM Parallel Applications Workshop, Alternatives To MPI (PAW-ATM)}, 2019, pp. 58--72.

\bibitem[Rocklin(2015)]{Rocklin2015DaskPC}
\BIBentryALTinterwordspacing
M.~Rocklin, ``Dask: Parallel computation with blocked algorithms and task scheduling,'' in \emph{SciPy}, 2015. [Online]. Available: \url{https://api.semanticscholar.org/CorpusID:63554230}
\BIBentrySTDinterwordspacing

\bibitem[Monti et~al.(2019)Monti, Frasca, Eynard, Mannion, and Bronstein]{montiFakeNewsDetection2019a}
F.~Monti, F.~Frasca, D.~Eynard, D.~Mannion, and M.~M. Bronstein, ``Fake {{News Detection}} on {{Social Media}} using {{Geometric Deep Learning}},'' 2019.

\bibitem[Kipf and Welling(2017)]{kipfSemiSupervisedClassificationGraph2017a}
T.~N. Kipf and M.~Welling, ``Semi-{{Supervised Classification}} with {{Graph Convolutional Networks}},'' 2017.

\bibitem[Kutalev and Lapina(2021)]{Kutalev2021StabilizingEW}
A.~Kutalev and A.~Lapina, ``Stabilizing elastic weight consolidation method in practical ml tasks and using weight importances for neural network pruning,'' \emph{ArXiv}, vol. abs/2109.10021, 2021.

\bibitem[Karakida et~al.(2019)Karakida, Akaho, and Amari]{karakida2019universal}
R.~Karakida, S.~Akaho, and S.-i. Amari, ``Universal statistics of fisher information in deep neural networks: Mean field approach,'' in \emph{The 22nd International Conference on Artificial Intelligence and Statistics}.\hskip 1em plus 0.5em minus 0.4em\relax PMLR, 2019, pp. 1032--1041.

\end{thebibliography}
\end{document}